\documentclass[a4paper,fleqn]{cas-dc}

\usepackage[numbers]{natbib}
\def\tsc#1{\csdef{#1}{\textsc{\lowercase{#1}}\xspace}}
\tsc{WGM}
\tsc{QE}
\tsc{EP}
\tsc{PMS}
\tsc{BEC}
\tsc{DE}
\begin{document}
\let\WriteBookmarks\relax
\def\floatpagepagefraction{1}
\def\textpagefraction{.001}

\shorttitle{Chasing Nomadic Worlds}

\shortauthors{Lingam et~al.}

\title [mode = title]{Chasing Nomadic Worlds: A New Class of Deep Space Missions}                      


\author[1,2]{Manasvi Lingam}

\cormark[1]

\ead{mlingam@fit.edu}

\affiliation[1]{organization={Department of Aerospace, Physics and Space Sciences, Florida Institute of Technology},
    city={Melbourne},
    postcode={32901}, 
     state={FL},
    country={USA}}

    \affiliation[2]{organization={Department of Physics and Institute for Fusion Studies, The University of Texas at Austin},
    city={Austin},
    postcode={78712}, 
     state={TX},
    country={USA}}

\author[3,4]{Andreas M. Hein}

 \affiliation[3]{organization={SnT, University of Luxembourg},
    addressline={29 Avenue J.F Kennedy},
    postcode={L-1855}, 
    country={Luxembourg}}

    \affiliation[4]{organization={Initiative for Interstellar Studies (i4is)},
    addressline={27/29 South Lambeth Road},
    city={London},
    postcode={SW8 1SZ}, 
    country={UK}}

\author[5]{T. Marshall Eubanks}

\affiliation[5]{organization={Space Initiatives Inc.},
    city={Newport},
    postcode={24128}, 
     state={VA},
    country={USA}}

\cortext[cor1]{Corresponding author}

\begin{abstract}
Nomadic worlds, i.e., objects not gravitationally bound to any star(s), are of great interest to planetary science and astrobiology. They have garnered attention recently due to constraints derived from microlensing surveys and the recent discovery of interstellar planetesimals. In this paper, we roughly estimate the prevalence of nomadic worlds with radii of $100\,\mathrm{km} \lesssim R \lesssim 10^4\,\mathrm{km}$. The cumulative number density $n_>\left(>R\right)$ appears to follow a heuristic power law given by $n_> \propto R^{-3}$. Therefore, smaller objects are probably much more numerous than larger rocky nomadic planets, and statistically more likely to have members relatively close to the inner Solar system. Our results suggest that tens to hundreds of planet-sized nomadic worlds might populate the spherical volume centered on Earth and circumscribed by Proxima Centauri, and may thus comprise closer interstellar targets than any planets bound to stars. For the first time, we systematically analyze the feasibility of exploring these unbounded objects via deep space missions. We investigate what near-future propulsion systems could allow us to reach nomadic worlds of radius $> R$ in a $50$-year flight timescale. Objects with $R \sim 100$ km are within the purview of multiple propulsion methods such as electric sails, laser electric propulsion, and solar sails. In contrast, nomadic worlds with $R \gtrsim 1000$ km are accessible by laser sails (and perhaps nuclear fusion), thereby underscoring their vast potential for deep space exploration.
\end{abstract}

\iffalse
\begin{highlights}
\item Cumulative number density of nomadic worlds estimated as function of the size range.
\item Detectability and distance of nearest nomadic worlds in given size range calculated.
\item Novel class of deep space missions to nomadic worlds delineated.
\item Utility of light sails and nuclear fusion for flybys of nomadic worlds underscored.
\end{highlights}
\fi

% Keywords

\maketitle

\section{Introduction}\label{SecIntro}
The notion that interstellar objects (ISOs) regularly pass through the Solar system, and that a fraction of them might even be captured, evinces a surprisingly rich and lengthy history, traceable to at least the writings of Pierre-Simon Laplace more than two centuries ago, reviewed in \cite{FJ83}. The field of ISOs was already quite active by the 1970s \cite{Wit72,Whip75,Sek76}, but it has received a timely boost thanks to the recent discovery of the interstellar planetesimals 1I/`Oumuamua and 2I/Borisov in 2017 \cite{MWM17} and 2019 \cite{JL19,GDR20}, respectively. Aside from this duo, interstellar dust grains and meteors have been analyzed, and their abundances constrained \cite{LBG00,MWB12,HSW19,ALAS,SL22}; note, however, that their sizes are many orders of magnitude smaller than the class of objects tackled hereafter. These two interstellar interlopers have substantially enhanced our knowledge of ISOs \cite{AMM22,JS22,SM23,FMMP}, such as the size distribution and number density of interstellar planetesimals, while raising new questions in their wake.

1I/`Oumuamua and 2I/Borisov are small (viz., sub-km size) objects with radii in the rough vicinity of $0.1$ km. Instead, if we turn our attention to much larger objects, the pioneering publications by \cite{MP91} and \cite{GL92} demonstrated that gravitational microlensing could enable the detection of planets, including those with long orbital periods or even potentially unbound to any star(s). Microlensing surveys have yielded a wealth of information concerning bound exoplanets \cite{SG12,YT18}, and gauged the distribution of unbound (i.e., free-floating) worlds \cite[e.g.,][]{SKB11,MUS17,DG18,BDG19}. Future surveys like the Nancy Grace Roman Space Telescope and Euclid are predicted to generate comprehensive data about this class of objects \cite{PBJ19,JPG20,BSP22}; the former may permit the discovery of worlds smaller than the Moon \cite{GZM21}.

The aforementioned observational avenues suggest that free-floating worlds, also known as rogue worlds, are prevalent in our Galaxy. Of particular interest are such objects belonging to the radius range of $\sim 100$ km to $\sim 10^4$ km. The upper bound is approximately the limit beyond which few rocky worlds exist based on exoplanet surveys \cite{LAR16,FPH17,FP18}. Moreover, roughly Earth-sized and larger rogue worlds have the capacity to host liquid oceans on their surface (up to Gyr timescales) from a combination of primordial and radiogenic heat; this hydrosphere can be present in the surface \cite{DJS99,VB11,Bad11,LL20,MHM,MPC21,RGE23} or subsurface \cite{AS11,LL19,MaLi20}. We exclude giant planet analogs, not only because they are rarer (as outlined later in Section \ref{SSecNumDens}) but also since their habitability potential is indeterminate, notwithstanding several publications on hypothetical pathways to abiogenesis on these worlds \cite{EJO64,Shap67,SS76,YPB17,ML19,SPG21}; in addition, they are not likely to be found close enough to the Earth for pursuing near term \textit{in situ} exploration.

The lower bound is comparable to the radius of Enceladus (although smaller by a factor of $\sim 2$), which manifestly fulfills most of the requirements for habitability \cite{PDM17,NAD20,CPG21,HGH22,MND22}. More crucially, objects with radii $\gtrsim 100$ km might retain habitable conditions -- liquid water more precisely, since it is conventionally viewed as a proxy for habitability \cite{SS13,MLAL} -- on timescales approaching $100$ Myr \cite{AM11}, namely, sufficient for ejection from one stellar system and entering the vicinity of another \cite{AS05,VNZ09,GLL18}. Habitability over this duration is rendered feasible in theory by heating derived from short-lived radionuclides \cite{AM11,LL20}, such as aluminium-26 \cite{MG15,LGB19}. Moving to larger objects of order $1000$ km, their habitability (i.e., liquid water sustaining) interval is boosted to Gyr timescales because of primordial heat and decay of long-lived radionuclides; several such worlds are documented or predicted in our Solar system and extrasolar planetary systems \cite{SS03,HSS06,AS11,CNS20,LL20,SJM21,CWB23}. 

On the other hand, conspicuously smaller objects with sizes $\lesssim 10$ km are unlikely to exhibit the preceding properties. Broadly speaking, as outlined in the earlier two paragraphs, our limits are compatible with our current understanding of habitable conditions and the objects that can host them. In our subsequent discussion, ISOs with radii $\gtrsim 100$ km will be dubbed \emph{nomadic worlds} (or nomads for short);\footnote{Incidentally, the proposed size threshold implies that this category of objects may often attain a crudely spherical shape \cite{DPL15}.} to put it another way, nomadic worlds are a specific subset of ISOs because the latter group also encompasses smaller objects (e.g., km-sized objects).

The habitability of nomadic worlds remains indeterminate, although a plethora of publications have identified possible avenues whereby these objects could be rendered suitable for hosting life \cite{HS58,EJO64,DJS99,RT01,DS07,AS11,VB11,GMK11,LL19,ML19,GS20,LL20,MaLi20,AGB21,MPC21,SMF21,MHM,LBM23,RGE23}. Hence, detecting and characterizing a nomadic world in the above size range would accord us a unique opportunity to gauge whether such abodes constitute viable habitats for life, evaluate our understanding of habitability, and determine whether prebiotic molecules and/or organisms may be transported across interstellar distances \cite{WM04,CW10,LL17,GLL18,GS20,SMF21}. 

Even setting aside the astrobiological potential of nomadic worlds, it is readily apparent that in-depth observations of even one such object could shed valuable light on, and revolutionize, many facets of planetary science. For instance, by extrapolating from previous publications on interstellar planetesimals, we may gain insights regarding the planetary systems and environments in which these objects formed \cite{FJ18,LL18,PTP18,MM18,LBM22,SRC22}, and the formation and dispersal/ejection of planetesimals and planets \cite{TRR17,RAV18,GPA19,PB19,RL19,WPL20,PAB21,PZ21,FMF22,MN22}. Furthermore, planets that form after the end of the stellar main sequence may not be observable in conventional stellar planetary systems, but ought to be present in the nomadic population \cite{EHL21}.  

Thus, viewed in totality, it is patently evident that the characterization of nomadic worlds could revolutionize planetary science and astrobiology, among other fields. However, one notable distinction must be spelled out at this stage. Limited characterization by Earth- and space-based telescopes is expected due to their likely substantial distances from the Earth, especially the larger nomadic worlds (see Section \ref{SSecDistISO}), and the transient observing windows likely for smaller objects. Therefore, the rationale for flyby missions to these objects is greatly strengthened, because such missions can mitigate the aforementioned two hurdles and permit detailed characterization. Even in the case of Pluto -- which is likely much closer to Earth than any large nomadic world -- the \emph{New Horizons} flyby yielded data about Pluto's atmosphere, surface, interior, and its satellites \cite{SGM18,SMG21} that were not obtainable from observations on Earth.

This paper introduces a new class of space missions, namely, to nomadic worlds in the Solar neighborhood, and explores their feasibility. The contents of the paper obey the following order. In Section \ref{SecDist}, rough estimates for the number density and characteristic distances of nomadic worlds of certain sizes are presented on the basis of the available empirical data. We follow this up by commenting on the prospects for detecting these objects in Section \ref{SecDetect}. Based on these two sections, we elucidate various propulsion technologies for reaching nomadic worlds in Section \ref{SecProp} and summarize our central findings in Section \ref{SecConc}.

\section{Typical minimum distances to nomadic worlds}\label{SecDist}
In this section, we estimate the number density of ISOs, with an emphasis on nomadic worlds with radii $\gtrsim 100$ km, as explained in Section \ref{SecIntro}. Next, this estimate is used to calculate the characteristic minimum distance to an ISO in the specified size range, and explore the ensuing ramifications. Owing to the scarcity of data on the ISO size distribution, we emphasize at the outset that our treatment is heuristic and the resulting estimates are ballpark figures.

\subsection{Heuristic number density of ISOs}\label{SSecNumDens}
Let us denote the (complementary cumulative) number density of ISOs with radii $> R$ by $n_>$. We shall use the following simplified expression henceforth:
\begin{equation}\label{NumDenISO}
    n_> \sim 0.1\,\mathrm{AU}^{-3}\,\left(\frac{R}{0.1\,\mathrm{km}}\right)^{-3}.
\end{equation}
A power-law exponent of roughly $-3$ has been adopted in (\ref{NumDenISO}) by drawing on available data regarding ISOs and interstellar dust grains \cite{LBG00,MWB12,ALAS,RL19,BLK20,JHK20,SL21,DJ22,SMM23}, as well as free-floating planets (addressed a couple of paragraphs later). The numerical values in this equation are based on constraints stemming from the discovery of 1I/`Oumuamua \cite{GWK17,JLR17,DTT18,PTP18,JHK20,EHL21,JS22,SMM23}. 

The exponent for the number density (\ref{NumDenISO}) is somewhat steeper than the prediction in \cite{SL19,ALAS}, who generated a power-law fit for interstellar meteoroids and meteors, but happens to be consistent with \cite{SL21}. We comment on the accuracy of extrapolating (\ref{NumDenISO}) to objects approaching planetary sizes below, as they are of immediate relevance to this paper. We reiterate that this expression is heuristic -- the primary objective of this work is not to quantify $n_>$ with high accuracy (which obviously calls for more data and modeling), but rather to employ it for analyzing propulsion methods well-suited to reaching different sizes of ISOs.

There are a couple of noteworthy caveats worth underscoring with respect to (\ref{NumDenISO}). First, as outlined in \cite{MTL09} and references therein (see also \cite{Oum19,AMM22}), the power-law exponent for the size distribution of ISOs is poorly constrained; as a result, it could comprise a series of power-law distributions, or not exhibit the power-law scaling altogether. Second, it is evident that (\ref{NumDenISO}) is dependent on $R$ and not the velocity of the ISOs. In other words, to formulate (\ref{NumDenISO}), we assumed that the majority of ISOs passing through the Solar system are endowed with broadly similar hyperbolic excess velocities, which, based on stellar velocity distributions, may represent a reasonable approximation to leading order \cite{EHL21,HSP21}.

In the preceding discussion, we constructed (\ref{NumDenISO}) from data of small objects. It is necessary to investigate how this expression stands up when extrapolated to larger objects. To examine this issue, we choose $R \approx R_\mathrm{Moon} \equiv 1737$ km (i.e., the radius of the Moon) and invoke (\ref{NumDenISO}), thereby obtaining $n_> \sim 1.7 \times 10^2$ pc$^{-3}$. On selecting a stellar number density of $\sim 0.1$ pc$^{-3}$ -- which is typical for the Solar neighborhood,\footnote{\url{http://www.pas.rochester.edu/~emamajek/memo_star_dens.html}} as well as the mid-plane of the Milky Way \cite{JB17} -- the number of objects with radii $> R_\mathrm{Moon}$ is $\sim 1.7 \times 10^3$ per star. Gravitational microlensing studies suggest that the corresponding value is comparable at $\sim 2 \times 10^3$ \cite{DG18,BDG19}, although some theoretical models indicate that this number may be higher by as much as $1$-$2$ orders of magnitude \cite{SBM12}.

On choosing $R \approx 0.75\,R_\oplus$, this radius can translate to a mass of about $0.33\,M_\oplus$ for rocky planets \citep{ZSJ16}. After substituting this choice of $R$ into (\ref{NumDenISO}), we arrive at $n_> \sim 8$ pc$^{-3}$. By employing the above stellar number density of $\sim 0.1$ pc$^{-3}$, we consequently obtain $\sim 80$ objects per star (with masses $\gtrsim 0.33\,M_\oplus$). A recent study by \cite{SKB23} based on the MOA-II microlensing survey of the Galactic bulge concluded that the corresponding value (for this mass range) is $\sim 20$ per star, i.e., the theoretical prediction is $\sim 4$ times higher. If we increase the threshold to $R \approx R_\oplus$, then we find $\sim 34$ objects per star after using (\ref{NumDenISO}) and repeating the analysis. This theoretical estimate is $\gtrsim 3$ times higher than the empirical constraints from the OGLE microlensing survey \citep{BSP22}. Thus, our heuristic model yields a modest overestimate of the number of nearly Earth-sized nomadic worlds that may exist per star in certain neighborhoods.

Next, moving on to giant planet analogs, we specify $R = R_J$ in (\ref{NumDenISO}), where $R_J$ stands for the radius of Jupiter, which duly yields $n_> \sim 2.6 \times 10^{-3}$ pc$^{-3}$. On using the stellar number density of $\sim 0.1$ pc$^{-3}$ from the preceding paragraphs, we obtain an abundance of $\sim 2.6 \times 10^{-2}$ super-Jovian planets per star. Microlensing surveys yielded constraints of $\sim 4.5 \times 10^{-2}$ \cite{MRB22}, $\lesssim 2.5 \times 10^{-1}$ \cite{MUS17}, $\lesssim 1$ \cite{CG17}, and $\lesssim 2$ \cite{SKB11} for the frequency of such worlds per star. A combination of numerical modeling and data from exoplanet surveys supports a frequency of $\sim 3 \times 10^{-2}$ super-Jovian planets per star; consult \cite{FH19}, \cite[Table 1]{MRB22} and \cite[Section 6]{SMJ22} for the details. Hence, our theoretical estimate, though simple, displays good agreement with the cited publications.

Lastly, based on the available data from microlensing surveys, \cite{GJH22} and \cite{SKB23} inferred that the mass distribution $N(M)$ of free-floating worlds per star appears to roughly obey $N(M) \propto M^{-2}$; in reality, the power-law exponent ranged between $-1.9$ and $-2.2$. A similar power-law exponent was derived for interstellar planetesimals in \cite{DJ22}. If we express the probability distribution in terms of $R$ via the canonical (albeit not exact) relationship $M \propto R^3$ and subsequently derive the (complementary) cumulative number distribution, it is found that the latter quantity is proportional to $R^{-3}$, which corresponds to the same power-law scaling as (\ref{NumDenISO}), thus serving as another consistency check.

It is evident, therefore, that our estimates derived from (\ref{NumDenISO}) are compatible with the data garnered from microlensing studies and other exoplanet surveys as well as computational models. In fact, it is conceivable that our ansatz (\ref{NumDenISO}) might even err on the side of caution in some ranges or models. Hence, in our subsequent analysis, we will draw on (\ref{NumDenISO}), but the previously highlighted caveats should be borne in mind.

\begin{figure}
\includegraphics[width=8.0cm]{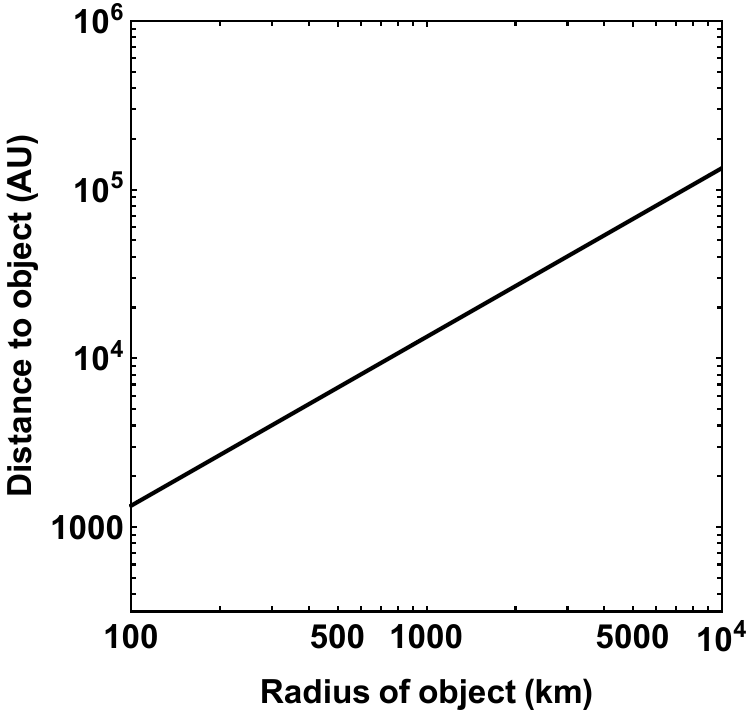} \\
\caption{The characteristic distance to a nomadic world (y-axis) with radius $> R$ (x-axis) based on (\ref{DISO}).}
\label{DistISO}
\end{figure}

\subsection{Distances to ISOs and other considerations}\label{SSecDistISO}
We are now in a position to calculate the characteristic distance $D_\mathrm{ISO}$ from Earth to the nearest ISO with radius $> R$. This quantity is crudely determined by means of
\begin{equation}
    \frac{4\pi}{3} n_> D_\mathrm{ISO}^3 \sim 1,
\end{equation}
which implies that one ISO (in the desired size range) must lie within a sphere of radius $D_\mathrm{ISO}$. We can now solve for $D_\mathrm{ISO}$ by invoking (\ref{NumDenISO}), thence obtaining
\begin{equation}\label{DISO}
    D_\mathrm{ISO} \sim 13.4\,\mathrm{AU}\,\left(\frac{R}{1\,\mathrm{km}}\right).
\end{equation}
In Figure \ref{DistISO}, we have plotted $D_\mathrm{ISO}$ as a function of the radius $R$. The latter variable is chosen to range from $100$ km to $10^4$ km, for reasons justified in Section \ref{SecIntro}.

At this juncture, a couple of implications of (\ref{NumDenISO}) and (\ref{DISO}) are worth spelling out explicitly.
\begin{itemize}
\item There is no ``hard'' cutoff for the minimum size when an object can be labeled a planet. Notwithstanding this caveat, if we select $R \approx 2440$ km (i.e., radius of Mercury) and $R \approx R_\oplus$ (an Earth-sized object), we end up with $D_\mathrm{ISO} \sim 0.16$ pc and $D_\mathrm{ISO} \sim 0.4$ pc, respectively. On comparing these values -- where Mercury and Earth represent optimistic and conservative choices for terrestrial planet sizes -- with the distance of $D_{PC} \approx 1.3$ pc to the Proxima Centauri planetary system,\footnote{At first, only a single planet (Proxima b) was detected around Proxima Centauri, but this star could harbor additional planets as hinted by some subsequent publications.} free-floating ``planets'' should be situated closer to Earth by a factor of $\sim 3$ - $8$ than the nearest known exoplanets around stars. As a result, the nearest ``planet'' to Earth may likely be a nomadic world, as hypothesized in \cite{MF90} and \cite{RT01}.
\item The number of free-floating worlds of radius $> R$ within distance $D_{PC}$ from Earth (denoted by $N_{PC}$), i.e., corresponding to the location of Proxima Centauri, can estimated using (\ref{NumDenISO}), yielding
\begin{eqnarray}
&& N_{PC} \approx \frac{4\pi}{3} n_> D_{PC}^3 \nonumber \\
&& \hspace{0.4in} \sim 8.1 \times 10^3\,\left(\frac{R}{1000\,\mathrm{km}}\right)^{-3}.
\end{eqnarray}
Thus, after substituting $R \approx 2440$ km and $R \approx R_\oplus$ in the same vein as the preceding point, we arrive at $N_{PC} \sim 5.6 \times 10^2$ and $N_{PC} \sim 31$, respectively. Therefore, in the same volume (centered on Earth) inhabited by the Proxima Centauri planetary system, it is possible that $\mathcal{O}(10)$ to $\mathcal{O}(100)$ free-floating planets exist; this number could even approach $10^3$ if the cutoff for $R$ is slightly lowered. This calculation underscores how much more plentiful nomadic planets might be (in the Solar neighborhood) relative to planets gravitationally bound to stars.
\end{itemize}

A crucial distinction concerning $D_\mathrm{ISO}$ warrants highlighting before proceeding further. While $D_\mathrm{ISO}$ embodies the typical distance where a nearby ISO (in the specified size range) may exist at any given time, it is not the same as: (1) the distance where an ISO of this size would be actually detected, and (2) the distance where a spacecraft would eventually intercept this ISO. Thus, it is necessary to address these two aspects, which is done below.

With regards to (1), it is plausible that the putative ISO would be detected close to its periapsis, which is not necessarily the same order as $D_\mathrm{ISO}$; the latter of which should therefore be taken as only a rough guide to the possible discovery distance. This is illustrated by the cases of 1I/`Oumuamua and 2I/Borisov. We predict for 1I/`Oumuamua-like objects with $R \sim 0.1$ km, a $D_\mathrm{ISO} \sim 1.3$ AU on employing (\ref{DISO}), whereas 1I/`Oumuamua was first detected at $\sim 1.2$ AU from the Sun and possesses a perihelion of $0.25$ AU \cite{MWM17,Oum19}. Likewise, 2I/Borisov was detected in 2019 at $\sim 3$ AU from the Sun, although there are ``precovery'' observations beginning in December 2018, when it was $\sim$7.8 AU from the Sun \cite{JHK20,YKB20}, both of which are similar to $D_\mathrm{ISO} \sim 5.4$ AU for $R \sim 0.4$ km \cite{HYF20}.

Next, let us turn our attention to (2). We have stated that the distance of the ISO from Earth or the Sun -- note that the Sun-ISO distance is close to the Earth-ISO distance if the latter is much larger than $1$ AU  -- is not the same as $D_\mathrm{ISO}$, although it might be comparable, as intimated by the prior paragraph based on the detection of 1I/`Oumuamua and 2I/Borisov. In determining the distance(s) from the Earth/Sun when the spacecraft is launched toward the ISO and intercepts it, there are multiple factors at play ranging from the hyperbolic trajectory parameters of the ISO (e.g., hyperbolic excess velocity, periapsis) to the time after detection that the spacecraft is launched, the maneuvers deployed, the propulsion schemes utilized, and so forth.

In our ensuing discussion, we will suppose that the spacecraft is launched toward the ISO shortly after its detection such that the distance $D_i$ between the spacecraft (at the time of launch) and the ISO is approximately $D_\mathrm{ISO}$. While this choice of $D_i \sim D_\mathrm{ISO}$ constitutes an idealization to some degree, it might not be unreasonable  provided that the detection occurs at distances $\gtrsim D_\mathrm{ISO}$ when the ISO is inbound into the Solar system and the spacecraft is designed and equipped beforehand for a quick launch thereafter (refer to \cite[Section 5.5.4]{SFJK}). Plausible mission architectures to intercept ISOs of sizes $\lesssim 1$-$10$ km based on different wait times, orbital maneuvers, and propulsion methods have been formulated by many authors \cite{SL18,HPE19,HHE20,HH21,HPH21,MCF21,SMH21,GFD22,HEH21,HHE22,HEL22,LH22,MDM22,SLM22,AH23,LDK23}.

\section{Detection of nomadic worlds}\label{SecDetect}
In this section, we will discuss some aspects of the detectability of ISOs, and clarify why our analysis is primarily oriented toward larger ISOs. We reiterate that our goal is to use these preliminary findings to pave the way for the engineering facets of our study, and a more exhaustive analysis of the prospects of detecting nomadic worlds is deferred to a companion publication.

The spectral flux density $S_\mathrm{ISO}$ associated with thermal emission from an ISO of radius $R$ and surface temperature $T_s$ can be roughly expressed as
\begin{equation}\label{SISOdef}
    S_\mathrm{ISO} \propto \frac{R^2 T_s^4}{D^2 \Delta \nu},
\end{equation}
where $D$ represents the distance of the ISO from Earth, and $\Delta \nu$ signifies the bandwidth associated with the thermal radiation. We will evaluate $S_\mathrm{ISO}$ near the Wien blackbody peak such that $\Delta \nu \sim \nu_\mathrm{peak} \propto T_s$. On substituting this relation into the above equation, we obtain
\begin{equation}\label{SISOint}
    S_\mathrm{ISO} \propto \frac{R^2 T_s^3}{D^2}.
\end{equation}
If we consider ISOs far away from the Sun, which is likely to be valid for large ISOs (with sizes $\gtrsim 100$ km) as seen from (\ref{DISO}) and Figure \ref{DistISO}, the internal temperature is chiefly governed by the geothermal heat, thereby yielding \cite{LL19},
\begin{equation}
    T_s \propto R^{0.33},
\end{equation}
for bodies with $R < R_\oplus$ composed of ice and/or rock. Lastly, we suppose that the ISO of interest is situated at $D \sim D_\mathrm{ISO}$ at the moment of detection, and then invoke (\ref{DISO}) to arrive at $D \propto R$. On substituting these relations into (\ref{SISOint}), we finally end up with the scaling
\begin{equation}\label{SISOfin}
     S_\mathrm{ISO} \propto R.
\end{equation}
The power-law exponent of $+1$ in (\ref{SISOfin}) must not be viewed as exact because of the multiple assumptions entering the derivation. Nevertheless, what this result does indicate is that larger ISOs are perhaps easier to detect than smaller ones even though the latter are more plentiful and potentially exist closer to Earth, as revealed by (\ref{NumDenISO}) and (\ref{DISO}). This qualitative trend is consistent with \cite{LL18}, where a similar procedure and outcome was delineated.

Let us explore another aspect of detectability, namely, the angular resolution of the ISO ($\theta_\mathrm{ISO}$) as seen from Earth, which is estimated to be
\begin{equation}\label{thetaISO}
    \theta_\mathrm{ISO} \approx \frac{R}{D} \sim 0.08\,\mathrm{mas} \left(\frac{R}{R_\oplus}\right)\left(\frac{D}{10^5\,\mathrm{AU}}\right)^{-1}.
\end{equation}
If we further suppose that $D \sim D_\mathrm{ISO}$ is typically applicable and harness (\ref{DISO}), then we find that (\ref{thetaISO}) is transformed into
\begin{equation}\label{thetaISOv2}
    \theta_\mathrm{ISO} \sim 0.1\,\mathrm{mas},
\end{equation}
where the RHS is obtained by invoking (\ref{DISO}). From the above expression, we see that $\theta_\mathrm{ISO}$ is independent of the size of the ISO; in actuality, it may exhibit a weak dependence because the power-law exponent in (\ref{NumDenISO}) is not exactly equal to $-3$. If (\ref{thetaISOv2}) is roughly valid, there is minimal penalty incurred by large ISOs compared to their smaller counterparts insofar as angular resolution is concerned.

Lastly, we examine the detectability of ISOs using current and future telescopes functioning primarily in optical wavelengths. The absolute magnitude $H$ can be calculated if the radius $R$ and the geometric albedo $\mathcal{A}$ of the putative ISO are known (see \cite{KKO07,MBC10}),\footnote{\url{https://cneos.jpl.nasa.gov/tools/ast_size_est.html}} via the expression
\begin{equation}
    \log_\mathrm{10}\left(\frac{R}{1\,\mathrm{km}}\right) \approx 2.82 - 0.5 \log_\mathrm{10} \mathcal{A} - 0.2 H,
\end{equation}
which upon rearrangement leads us to
\begin{equation}\label{AbsMagInt}
    H \approx 14.1 - 2.5 \log_\mathrm{10}\left[\left(\frac{R}{1\,\mathrm{km}}\right)^2 \mathcal{A} \right].
\end{equation}
The apparent magnitude $m$ is related to the absolute magnitude through the relation \cite{SL18},
\begin{equation}\label{AppMag}
    m = H + 2.5 \log_\mathrm{10}\left(\frac{D_\mathrm{IE}^2 D_\mathrm{IS}^2}{P(\alpha) D_\mathrm{SE}^4}\right),
\end{equation}
where $D_\mathrm{IE}$, $D_\mathrm{IS}$, and $D_\mathrm{SE} \approx 1$ AU represent the ISO-Earth (presuming that the observer is on Earth), ISO-Sun, and Earth-Sun distances, respectively, while $\alpha$ and $P(\alpha)$ are the phase angle and phase integral. Since we are interested in ISOs of sizes $\gtrsim \mathcal{O}(100)$ km, it is likely that $D_\mathrm{IE} \sim D_\mathrm{IS} \gg D_\mathrm{SE}$ as implicitly outlined in Section \ref{SecDist}. Based on the assumptions adumbrated in Section \ref{SecDist}, we furthermore specify $D_\mathrm{IE} \sim D_\mathrm{IS} \sim  D_\mathrm{ISO}$. The phase angle is defined as
\begin{equation}
    \cos \alpha = \frac{D_\mathrm{IE}^2 + D_\mathrm{IS}^2 - D_\mathrm{SE}^2}{2 D_\mathrm{IE} D_\mathrm{IS}},
\end{equation}
and it simplifies to $\cos \alpha \rightarrow 1$ using the expressions in the preceding paragraph, which translates to $\alpha \rightarrow 0$. The phase integral is computed from
\begin{equation}
   P(\alpha) = \frac{2}{3}\left[\left(1 - \frac{\alpha}{\pi}\right)\cos \alpha + \frac{1}{\pi} \sin \alpha \right], 
\end{equation}
and it reduces to $P(\alpha) \sim 2/3$ when we consider the above case of $\alpha \rightarrow 0$. On using this value in (\ref{AppMag}) along with the aforementioned choices for the distances, the apparent magnitude is expressible as
\begin{equation}\label{AppMagInt}
    m \approx H + 11.7 + 2.5 \log_\mathrm{10}\left[\left(\frac{R}{1\,\mathrm{km}}\right)^4  \right],
\end{equation}
after employing (\ref{DISO}). Lastly, on combining (\ref{AppMagInt}) with (\ref{AbsMagInt}), we duly end up with
\begin{equation}
    m \approx 25.8 + 2.5 \log_\mathrm{10}\left[\left(\frac{R}{1\,\mathrm{km}}\right)^2 \mathcal{A}^{-1} \right].
\end{equation}
The Vera C. Rubin Observatory is reported to have a sensitivity of $m \sim 24.5$ and final survey depth of $m \sim 27.5$ (over multiple visits) for a certain exposure time, as documented in \cite{IKT19}. With a much lower field-of-view relative to the Vera C. Rubin Observatory, the Hyper Suprime-Cam SSP Survey is capable of reaching a sensitivity of $m \sim 26$-$27$ \cite{AAA18,AAB18}. Hence, if we were to input $R \sim 1000$ km and $\mathcal{A} \sim 0.1$, this simplified model would predict $m \approx 43.3$, which is clearly much fainter than either of these two telescope thresholds.

However, several crucial caveats and clarifications can modify, and possibly bypass, the preceding conclusion(s). First, we have chosen to adopt $D_\mathrm{ISO}$ because it represents the typical distance to an ISO (of a particular size range) at any given moment in time. Alternatively, if the ISO were to be detected close to perihelion and this quantity is substantially smaller than $D_\mathrm{ISO}$, then the corresponding value of $m$ could become significantly lowered as per (\ref{AppMag}), consequently making detection by the Vera C. Rubin Observatory or the Hyper Suprime-Cam SSP Survey potentially viable. 

Second, the preceding analysis focused exclusively on the Vera C. Rubin Observatory and the Hyper Suprime-Cam SSP Survey, both of which are designed to probe only certain wavelength ranges (e.g., optical).\footnote{\url{https://www.lsst.org/scientists/keynumbers}} Instead, if we consider the far-infrared and seek to detect thermal emission from ISOs, it was demonstrated at the beginning of this section that the spectral flux density scales \emph{positively} with $R$, as illustrated by (\ref{SISOfin}). Therefore, as demonstrated in \cite[Section 3]{AS11} and \cite{LL19,LL21}, a roughly Earth-sized nomad may be detectable up to distances of $\sim 10^3-10^4$ AU (under optimal circumstances) by upcoming telescopes such as the upcoming Fred Young Submillimeter Telescope.\footnote{\url{https://www.ccatobservatory.org/index.cfm}}

Third, and perhaps most importantly, we have not tackled all the salient avenues for discovering nomadic worlds. We will briefly delineate one of them herein, namely, detecting nomadic worlds via stellar occultations; others will be thoroughly explored in a forthcoming publication, where the quantitative calculations will also be elucidated. A simple criterion that ought to permit stellar occultations by nomadic worlds is $\theta_\mathrm{ISO} > \theta_\star$, where $\theta_\star$ is the angular resolution of the star. In qualitative terms, this condition implies that the nomadic world can block the light from a distant star, thereby serving as a means of discerning the former. On simplifying this criterion, we arrive at
\begin{equation}
    d_\star \gtrsim 46.5\,\mathrm{pc}\left(\frac{R_\star}{R_\odot}\right) \left(\frac{\theta_\mathrm{ISO}}{0.1\,\mathrm{mas}}\right)^{-1},
\end{equation}
where $R_\star$ denotes the stellar radius, and $d_\star$ is the distance of this star from Earth. Next, we emphasize that $\theta_\mathrm{ISO}$ may turn out to have a minimal dependence on $R$, as explained in the paragraphs subsequent to (\ref{thetaISO}). Hence, in light of the possibly weak dependence of $\theta_\mathrm{ISO}$ and $d_\star$ (by extension) on the radius $R$, some of the results obtained for smaller ISOs with sizes $\lesssim 10$ km might be applicable to larger nomadic worlds. Numerical modeling by \cite{SL20} suggests that $\sim 1$ ISO per year could be detected by an all-sky survey monitoring $\sim 7 \times 10^6$ stars with a magnitude of $\lesssim 12.5$.

Distant objects with only a single isolated occultation (or any other type of) observation will not be suitable targets for \textit{in situ} exploration, as repeated measurements are needed; as stated earlier, various detection techniques of nomadic worlds will be covered in a companion publication. Occultation arrays will, however, have some advantages compared to more conventional optical observations (summarized in \cite{ABG15,MRL19}). An occultation of a previously unknown object by a suitable occulation array (i.e., one with multiple telescopes distributed over a region of order $100$ or $1000$ km) is expected to provide a direct estimate of its angular position in the sky and also of its physical velocity relative to the Earth, by measuring the speed of its shadow on the ground \cite{RGD08}. The benefits of a large-scale occultation array for identifying objects in the outer Solar system at distances of $\gtrsim 100$ AU are chronicled in \cite{SL20,MRL19}.

Since the location of the object in the sky is determined in an occultation, and the ephemerides of the Earth’s orbit and rotation are also well determined, for an occulation discovery, it will be possible to determine the transverse velocities of any detected objects relative to the solar system barycenter (SSB). For objects at distances of order $1000$ AU, it should be possible to determine angular positions to order $10^{-9}$ rad and velocities to order $10$ m/s. The former ensues from the preceding discussion of the $\theta$'s and (\ref{thetaISOv2}); this positional variation is significantly lower than the distinctly conservative value considered in Section \ref{SSecConFly}.

At the distance $D_\mathrm{ISO}$ from the Sun, the circular orbit velocity is $\sqrt{G M_\odot/D_\mathrm{ISO}} \sim 1\,\mathrm{km/s} \,\left(D_\mathrm{ISO}/1000\,\mathrm{AU}\right)^{-1/2}$. This relation means that it should be possible to distinguish immediately between bound objects (with orbital velocities of a few km/s or less) and unbound ISOs, with typical SSB velocities of several tens of km/s \cite{HSW19,SL20,EHL21,SLM22}. The dynamical observables (two components each of position and velocity) that can be computed from a single occultation are not enough, of course, to fully determine the orbit of the newly discovered object. Achieving this would require at least either another occultation or detection by a deep survey.

The initial occultation yields a physical velocity on the sky, while a subsequent occultation or direct observation would furnish a angular velocity, such that a comparison of these velocities would constrain the distance to the occulter; a preliminary orbit may be obtained even if the following observations are performed within a day or two of the initial detection of the object. Once two or more observations are acquired of this occultation-based discovery, the orbit could be refined by increasing the velocity accuracy through a series of occultations, thus paving the way for \emph{in situ} exploration of the nomadic world.

In the remainder of this paper, we will investigate the central question of what type(s) of propulsion systems may be feasible for intercepting a given nomadic world in the future in the serendipitous event that it would be discovered, and its location ascertained, through a suitable avenue.

\section{Propulsion systems for reaching nomadic worlds}\label{SecProp}
In this section, we will explore what types of near-future propulsion methods could permit us to reach nomadic worlds in a specific size range.\footnote{The word ``near-future'' is inherently ambiguous, but is usually taken to comprise a few decades in our treatment.} 

\subsection{Desirable terminal speeds for propulsion systems}\label{SSecTermS}

\begin{figure}
\includegraphics[width=8.0cm]{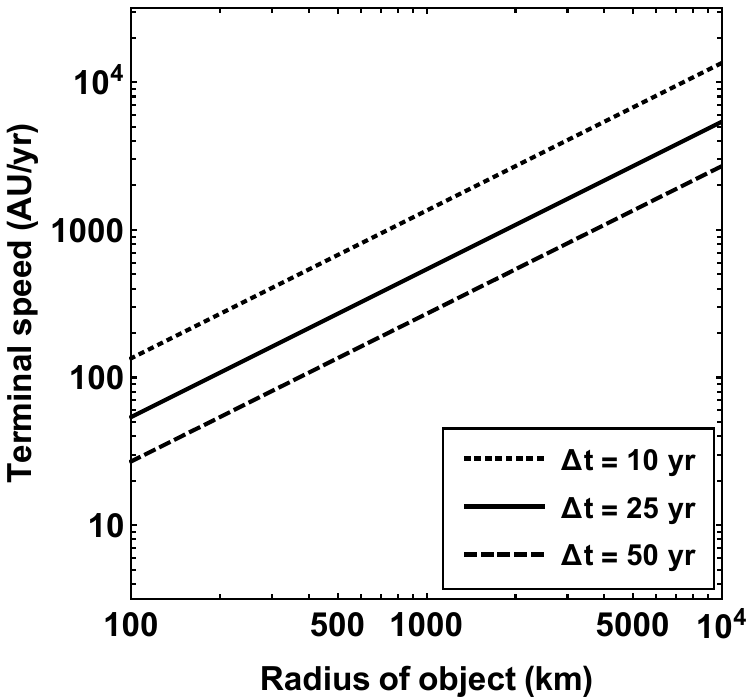} \\
\caption{The desired terminal speed associated with a propulsion system (y-axis) to reach a nomadic world with radius $> R$ (x-axis) for three different choices of the flight duration: $\Delta t = 10$ yr, $\Delta t = 25$ yr, and $\Delta t = 50$ yr; this plot is generated by invoking (\ref{vtapprox}).}
\label{TermSpeed}
\end{figure}

As measured from Earth's frame of reference, one could compute the flight duration to the nomadic world ($\Delta t$) via solving the complex ODE:
\begin{equation}\label{EOMISO}
d{\bf r} = {\bf v}_\mathrm{rel}\,dt \quad \Rightarrow \quad dr = v_\mathrm{rel}\,dt,
\end{equation}
where ${\bf v}_\mathrm{rel} = {\bf v}_\mathrm{sc} - {\bf v}_\mathrm{ISO}$ is the relative velocity, $r = |{\bf r}|$, and $v_\mathrm{rel} = |{\bf v}_\mathrm{rel}|$. Both the spacecraft velocity (${\bf v}_\mathrm{sc}$) and the ISO velocity (${\bf v}_\mathrm{ISO}$) explicitly depend on ${\bf r}$, thereby making it difficult to solve (\ref{EOMISO}) exactly. However, two simplifications can render the problem tractable without much loss of generality; a more comprehensive analysis of $\Delta t$ will constitute the subject of future work.
\begin{enumerate}
\item As stated above, ${\bf v}_\mathrm{sc}$ is actually a function of ${\bf r}$ because the spacecraft is accelerated to a final (i.e., terminal) speed $v_t$ over some distance $D_t$. However, as long as $D_t$ is much smaller than $D_i \sim D_\mathrm{ISO}$, we can roughly assume that the spacecraft travels at the (constant) speed $v_t$ for the majority of its journey, i.e., the trajectory is largely ballistic. As discussed later, for the propulsion technologies considered henceforth, $v_t$ is achievable over distances of $\lesssim 100$ AU, while $D_\mathrm{ISO}$ for nomadic worlds is $\gtrsim 1000$ AU, as seen from inspecting Figure \ref{DistISO}. Hence, as per this result, we may prescribe $|{\bf v}_\mathrm{sc}| \sim v_t$ during the voyage.
\item The hyperbolic excess velocities ($v_\infty$) of most ISOs are ostensibly few/several tens of km/s \cite[see, e.g.,][]{HSW19,SL20,EHL21,SLM22}. As long as the perihelia of these ISOs are not close to the Sun, it is reasonable to suppose that ${\bf v}_\mathrm{ISO}$ is on the order of $v_\infty$ throughout. In contrast, the terminal speeds of the propulsion systems that we tackle hereafter are $\gtrsim 100$ km/s. Thus, to leading order, we may approximate ${\bf v}_\mathrm{rel}$ by $v_t$ since the latter is expected to be sufficiently large (e.g., a few times higher) compared to $v_\infty$ and ${\bf v}_\mathrm{ISO}$ for most ISOs.
\end{enumerate}
With these two simplifications, we see that (\ref{EOMISO}) is transformed into the much simpler $dr = v_t\,dt$, which is readily integrated (after imposing integration limits) to yield
\begin{equation}
    \Delta t \sim \frac{D_\mathrm{ISO}}{v_t}.
\end{equation}
On solving this heuristic equation for $v_t$, we obtain
\begin{equation}\label{vtapprox}
    v_t \sim 0.27\,\mathrm{AU/yr}\,\left(\frac{R}{1\,\mathrm{km}}\right)\left(\frac{\Delta t}{50\,\mathrm{yr}}\right)^{-1}.
\end{equation}
The immediate advantage of (\ref{vtapprox}) is that it enables us to approximately gauge the desirable terminal speed to reach an ISO (of particular size range) in a given time interval $\Delta t$. Once this quantity is estimated, it can then be employed to determine suitable propulsion systems capable of attaining the requisite values of $v_t$.

In Figure \ref{TermSpeed}, the terminal speed $v_t$ is plotted as a function of $R$ for three different choices of $\Delta t = 10$ yr (i.e., rapid transit), $\Delta t = 25$ yr, and $\Delta t = 50$ yr.\footnote{We do not address $\Delta t = 10$ yr here onward, as it translates to significantly high speeds and energetic requirements; the latter scales as the square of the former.} The last two values are equivalent to the typical length of one and two human generations, respectively. Even though $\Delta t = 50$ yr may appear to be excessively long, it is worth recognizing that the two \emph{Voyager} spacecraft were launched $45$ years ago,\footnote{\url{https://voyager.jpl.nasa.gov/mission/}} and continue to collect and transmit data of scientific importance. Moreover, proposed future missions such as the Interstellar Probe concept \cite{KSR21,MWG22} and flyby missions to the putative Planet 9 \cite{HLH22} entail flight times of $\gtrsim 50$ years.

\subsection{Analysis of propulsion technologies}\label{SSecPropAnlys}
On the basis of (\ref{vtapprox}) and Figure \ref{TermSpeed}, we consider various propulsion technologies and estimate the range of nomadic worlds that they could survey.

A few clarifications should be borne in mind. First, our list is not exhaustive, in the sense that we do not cover every single near-future technology capable of attaining final speeds of order $100$ km/s (or more). Second, in a similar vein, we exclude propulsion systems predicated on the likes of antimatter as well as the interstellar ramjet, even though such spacecraft can reach the desired speeds (see \cite{Long11,Long22} and \cite[Chapter 10]{LL21}). Our rationale for omitting them stems from the fact that practical implementation of these propulsion systems is arguably unrealistic in the near future even from a purely technical standpoint \cite{KYM10}. 

Last, we mostly bypass multimode propulsion systems despite their advantages \cite{RLM20}, such as the combined electric and magnetic sail described in \cite{PH16}, because of the attendant complexities of identifying and designing appropriate combinations. To put it another way, some of the technologies we shall delineate -- like magnetoplasmadynamic thrusters on the one hand and nuclear/laser electric propulsion on the other -- can be consolidated to construct more efficient and/or swifter propulsion systems.

\subsubsection{Solar sails}\label{SSSecSolSail}
The concept of using radiation pressure to propel spacecraft (i.e., light sails) has a long history: the first modern iterations were published nearly one century ago \cite[e.g.,][]{Zand24}. In the case of solar sails, the radiation is derived from the Sun, obviating the need for an external power source. This propulsion method has witnessed extensive research and accompanying progress \cite{McIn04,JYM11,Vul12,GM19}. Successful prototypes of solar sails have been demonstrated such as \emph{IKAROS} \cite{MSF11,TMF13} and \emph{LightSail 2} \cite{SBB21}. 

We caution, however, that the technology readiness level (TRL) of solar sails has a current maximum of $7$ to $9$ only in optimistic circumstances \cite{GM19,TGF23}. One prominent caveat to note is that these ratings were assigned for dramatically different applications and environments. In contrast, if one considers the comparatively ambitious mission parameters (e.g., terminal speeds) addressed herein, it seems plausible that the TRL of such solar sails would be lowered to $3$-$4$ \cite{KBF20}, after recognizing that the TRL corresponds to that of the least-developed subsystem.

Ample research has been conducted on materials suitable for light sails in recent times \cite{ADI18,DMT21}. If these materials are successfully deployed, achieving $v_t \sim 20$ AU/yr seems feasible for solar sails launched near the Earth \cite{ML20}, in the absence of additional orbital maneuvers that further boost the speed \cite{JDB29}. At closer distances to the Sun, the terminal speed is enhanced accordingly by a few times \cite{AKM19}. If an aerographite sail were launched at merely $0.04$ AU from the Sun, the final speed might become as high as $\sim 1000$ AU/yr for small payloads of order $10^{-3}$ kg \cite{HAH20};\footnote{Note that weakly relativistic speeds are possible near high-energy astrophysical objects \cite{LM20}, but such exotic settings are manifestly inaccessible to humans.} unforeseen engineering issues due to thermal effects could, however, arise in this scenario.

With existing solar sail materials, a kg-sized probe can achieve $v_t \sim 6$ AU/yr and consequently target nearby ISOs at distances of order $10$ AU \cite{GFD22}; the associated acceleration distance is much smaller than $D_\mathrm{ISO}$ for nomadic worlds. Combining solar sails with a rocket engine and taking advantage of the Oberth effect has the potential to engender $v_t \sim 30$ AU/yr \cite{BJ21}. A realistic solar sail mission with spacecraft mass of $\sim 50$ kg to the Solar Gravitational Lens (SGL) at $\gtrsim 550$ AU may attain $v_t \sim 20$ AU/yr (or even higher values by a factor of about $2$) after performing appropriate orbital maneuvers in proximity to the Sun \cite{FGT21,HRP22,TGF23}, which builds on the concept espoused by \cite{JDB29}.

As our focus is on near-future technologies, we will opt for $v_t \sim 20$ AU/yr instead of lower or higher speeds, both of which are tenable. On substituting this terminal speed into (\ref{vtapprox}) and selecting a mission duration of $\Delta t = 50$ yr, we obtain $R \sim 75$ km. As the size range primarily considered in this paper is $10^2\,\mathrm{km} \lesssim R \lesssim 10^4\,\mathrm{km}$, the above value of $R$ falls slightly below the lower bound. In light of the uncertainties and approximations in our analysis, it is plausible that solar sails can be employed to survey nomadic worlds, but only those on the smaller end of the spectrum, i.e., around $100$ km in radius. 

\subsubsection{Laser sails}

\begin{figure}
\includegraphics[width=8.0cm]{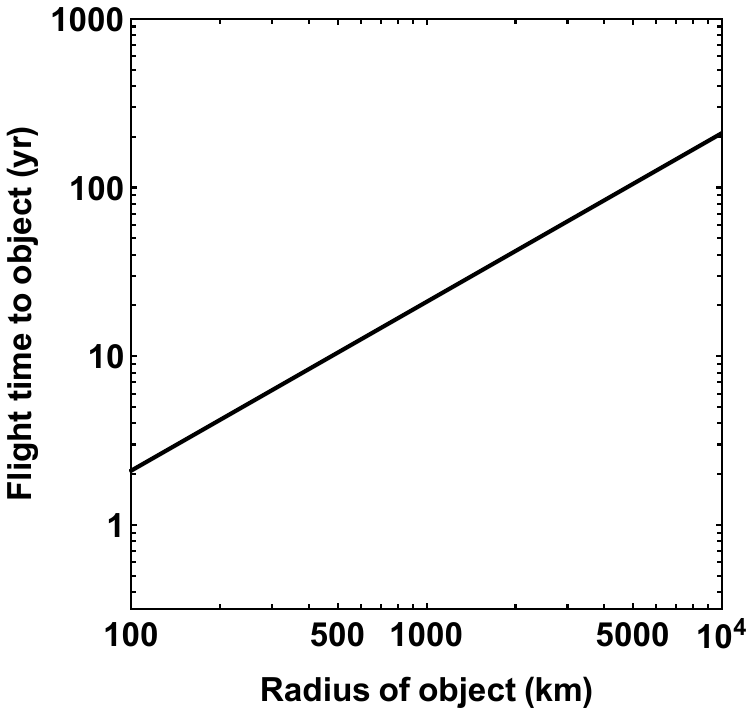} \\
\caption{The typical flight time expected to reach a nomadic world (y-axis) with radius $> R$ (x-axis) by utilizing a laser sail traveling at $1\%$ the speed of light ($3000$ km/s); the plot is generated from (\ref{DelTreq}).}
\label{LaserFlight}
\end{figure}

The underlying principle of laser sails is essentially identical to that of solar sails. The chief difference is that the former relies on lasers to supply the requisite radiation pressure (and power) instead of solar radiation associated with the latter. Laser sails were formulated and modeled less than a decade after the construction of the first working lasers \cite{Marx66,JR67}, and many publications have subsequently explored their feasibility \cite[e.g.,][]{Forw84,AM14,Lub16,KP18,PL22a}. It is worth noting that the ballistic assumption outlined in Section \ref{SSecTermS} is generally valid for laser sails \cite{MLM22}.

Laser sails have attracted renewed interest after the inception of the \emph{Breakthrough Starshot Initiative} in 2016,\footnote{\url{https://breakthroughinitiatives.org/initiative/3}} which aims to send a laser sail with payload of $\sim 10^{-3}$ kg to the nearest star(s) at $\sim 0.2\,c$ and flight time of $\sim 20$ years \cite{PD17,KP18}; earlier designs of lightweight spacecraft powered by beamed energy include \cite{RLF85} and \cite{HLF17}. Many engineering challenges confront this endeavor, some of which are reviewed in \cite{MLM,LL21,PLB21,WGS21}. As opposed to attaining speeds of order $0.1\,c$, several recent laser sail designs have emphasized the prospects of reaching terminal speeds of $\sim 1\%$ or $\sim 0.1\%$ the speed of light. While these proposals would lead to longer flight durations for a given distance, they have the advantages of enabling lower power requirements and/or greater payloads \cite{KP22,HD22}.

A number of recent mission designs have adopted the fiducial value of $300$ km/s ($10^{-3}\,c$) for deep space exploration; targets in the outer Solar system that can be surveyed by such laser sails include the heliopause, SGL, and Planet 9 \cite{TML20,HLH22}. At this speed, \cite{TML20} suggested that payloads of $\sim 1$ kg to $\sim 100$ kg would translate to power input in the range of $\sim 3-30$ GW. The detailed cost-optimization model of \cite[Table 2]{KP22} necessitates a peak power of $2.5$ GW to accelerate a $10$ kg payload to $300$ km/s; the capital expenditure and operating expenditure (per mission) are respectively predicted to be $\$610$ M and $\$58$ M. \cite{KP22} also calculated the power and cost requirements for certain payload masses at higher speeds of $\sim 0.01\,c$ and $\sim 0.1\,c$.

To determine what terminal speeds are needed by laser sails to study nomadic worlds, we consider two distinctive cases. First, we consider $\Delta t = 25$ yr and $R = 10^4$ km, which corresponds to reaching a world larger than Earth in a relatively short timescale. On substituting these values in (\ref{vtapprox}), we obtain $v_t \sim 5.4 \times 10^3$ AU/yr. Although this speed of $\sim 0.1\,c$ is comfortably above what can be achieved by solar sails (see \cite{HAH20}), laser sails analogous to \emph{Breakthrough Starshot} can achieve the desired speed, albeit with the drawback of high energy expenditures and/or low-mass payloads.

Now, let us proceed to contemplate the other limit with $\Delta t = 50$ yr and $R = 100$ km, which amounts to reaching a small nomadic world (i.e., radius lower than Enceladus) over a longer timescale. By employing these choices in (\ref{vtapprox}), we end up with $v_t \sim 27$ AU/yr. As remarked in the last paragraph of Section \ref{SSSecSolSail}, this speed is realizable by solar sails launched at Earth's location. By the same token, this threshold is readily achievable by light sails, as the desired speed is a modest $\sim 130$ km/s.

Given that laser sails can theoretically achieve speeds of $\sim 0.001\,c$ to $\sim 0.1\,c$, we will take the geometric mean of $\sim 0.01\,c$ as a fiducial value, and substitute this choice of $v_t$ in (\ref{vtapprox}). On doing so, we can estimate the approximate flight duration needed in order to reach a nomadic world of radius $> R$. After simplification, we arrive at
\begin{equation}\label{DelTreq}
\Delta t \sim 2.1 \times 10^{-2}\,\mathrm{yr}\,\left(\frac{R}{1\,\mathrm{km}}\right).
\end{equation}
As per this formula, even a nomadic world of radius $> 1000$ km might be reached by a laser sail traveling at $\sim 0.01\,c$ in $\sim 20$ years. We have plotted (\ref{DelTreq}) in Figure \ref{LaserFlight} for the size range of nomadic worlds introduced earlier.

Lastly, we remark that, in lieu of beamed photons, one could deploy particle beams for instantiating momentum transfer and accelerating the spacecraft. Such proposals have a fairly long history \cite{CES80,GAL04}. In conjunction with other propulsion methods such as light sails and magnetic sails \cite{GYL22,LM22}, or otherwise \cite{GDN93}, spacecraft propelled by particle beams are capable of attaining final speeds of order $0.01\,c$. However, because no actual prototypes exist and the number of theoretical studies are fewer, we do not delve deeper into this option.

\subsubsection{Electric sails and magnetic sails}\label{SSSecESail}
Electric sails \cite{Jan04} and magnetic sails \cite{AZ90,ZA91}, share certain similarities since they both depend on momentum exchange with the solar (or stellar) wind to attain fairly high velocities, with the added benefit of doing away with onboard fuel. Furthermore, electric sails and magnetic sails each have a well-defined upper limit corresponding to the typical speed of the solar/stellar wind, i.e., around $400$ km/s, as higher spacecraft speeds would lead to a net deceleration instead.\footnote{However, modifications of these designs may permit speeds higher that of the wind to be attained \citep{LHG22}.} Reviews of the magnetic and electric sail concepts can be found in \cite{HD18} and \cite{BNQ22}.

Yet, there are a couple of noteworthy differences between them that merit mention. First, prototypes of electric sails have been demonstrated in the past decade: the nanosatellites ESTCube-1 \cite{SPK15} and Aalto-1.\footnote{\url{https://www.aalto.fi/en/spacecraft/mission-and-science-results}} In contrast, no such prototypes of magnetic sails appear to exist, reducing its TRL accordingly. Second, on a related note, the classical version(s) of magnetic sails are reliant on robust superconducting materials, which introduces an extra complication \cite[e.g.,][]{GV94}. Hence, based on these two reasons, we shall focus primarily on electric sails henceforth. Before moving on, however, we remark that propulsion systems inspired by the classic magnetic sail could be accelerated to speeds approaching $100$ km/s \cite{WSZG00}.

By performing numerical modeling, \cite{JP19} concluded that the electric sail can accelerate a $500$ kg spacecraft to a terminal speed of $\sim 12$ AU/yr, thence proving appropriate for deep space missions. In theory, higher speeds of $\sim 25$ AU/yr may be attainable by realistic electric sail designs \cite[Section 6]{JS07}; see also \cite{PJ09} and \cite{JTP10}. The distances required for reaching these speeds are comfortably smaller than $D_\mathrm{ISO}$ \cite{JTP10}, which preserves consistency with the conditions in Section \ref{SSecTermS}. If the electric sail is launched in close proximity to the Sun, the ensuing terminal speeds can be boosted by a few times; the upper bound of $\sim 80$ AU/yr, as mentioned above, is set by the solar wind.

If we specify $v_t \sim 25$ AU/yr along with flight time of $\Delta t = 50$ yr in (\ref{vtapprox}), we obtain $R \sim 93$ km. Therefore, it is plausible that electric sails could enable us to study nomadic worlds of radius $\sim 100$ km on a timescale of $50$ years. Even though electric (and magnetic) sails do not seem suitable for reaching nomadic worlds of order $1000$ km, they may nevertheless represent realistic propulsion systems for surveying smaller worlds.

\subsubsection{Electric propulsion}\label{SSSecEP}
There are manifold mechanisms comprising electric and electromagnetic propulsion, owing to which they cannot be readily grouped under a single umbrella; reviews of this broad subject are furnished in \cite{EC09}, \cite{SM16} and \cite{LXM20}.

The first system that we wish to highlight is magnetoplasmadynamic (MPD) thrusters, which may be powered by various sources, including those discussed in the following paragraphs. Gaseous propellant ionized via electron impacts (thus forming plasma) is subjected to the Lorentz force, which duly accelerates the plasma to substantial speeds in MPD thrusters. MPD thrusters can not only generate high exhaust speeds approaching $v_{ex} \sim 100$ km/s, but also concomitantly provide high thrust densities up to $\sim 10^5$ N/m$^2$ (see Section 3.2 of \cite{SM16}) at high efficiencies \cite{BBH19}. We will hereafter invoke the heuristic that $v_t \lesssim 3 v_{ex}$ for a multi-stage rocket and reasonable payload mass ratio of $\sim 0.01$-$0.1$ \cite[pg. 57]{UW19}. Hence, we obtain $v_t \lesssim 300$ km/s on utilizing the information outlined in this paragraph.

Prototypes of MPD thrusters have already been validated,\footnote{\url{https://nap.nationalacademies.org/read/25977/chapter/5}} albeit not with the aforementioned parameters, thereby establishing proof-of-concept. The chief downside is that their efficient functioning is achieved at the cost of relatively significant power requirements. If this aspect is low priority, then MPD thrusters could achieve $v_t \lesssim 63$ AU/yr (motivated in the preceding paragraph) over distances much shorter than $D_\mathrm{ISO}$ (consult the criteria in Section \ref{SSecTermS}), thereby capable of reaching nomadic worlds with radii $\lesssim 230$ km in $50$ years. Some of the delineated pros and cons are ostensibly attributable to the Variable Specific Impulse Magnetoplasma Rocket (see \cite{CD00,CC09}).

Propulsion systems that integrate electric propulsion with appropriate energy sources have the capacity to attain high final speeds. One such major example is nuclear electric propulsion (NEP), which could theoretically impart velocities of $\mathcal{O}(100)$ km/s to the spacecraft \cite{NPL01,ASP20}. Although NEP currently possesses some limitations -- for example, some of its subsystems apparently involve technologically immature components \cite{SPW21} -- in the long term, this propulsion scheme ought to be rendered viable \cite{JRC20,OHK21},\footnote{\url{https://www.nasa.gov/press-release/demonstration-proves-nuclear-fission-system-can-provide-space-exploration-power}} given that this technology is already assigned TRL-5 as per some sources \cite{MPG20,PMC22}. Even in the near term, this propulsion system might enable rendezvous missions to ISOs at relatively close distances of $\lesssim 100$ AU \cite{LDK23}.

Next, we will turn our attention toward the avenue of laser electric propulsion (LEP). Akin to laser sails, LEP necessitates directed energy (electromagnetic radiation in this context) beamed to the spacecraft. This energy can be harnessed by onboard photovoltaics and subsequently employed to power ion thrusters \cite{DLN76,BPA18}.\footnote{We shall not address the partly related concept of laser ablation propulsion \cite{JTK03,PBB10}, which could engender exhaust velocities of $\lesssim 50$ km/s \cite{PBL10,OGV,LBM}.} LEP might facilitate the attainment of terminal speeds as high as $v_t \sim 40$ AU/yr \cite{BPA18}. Other in-depth mission designs predicated on LEP have yielded lower speeds of order $10$ AU/yr \cite{SPW21}. The viability of LEP is difficult to encapsulate because the various subsystems have different TRLs \cite{SPW21}. On plugging $v_t \sim 40$ AU/yr and $\Delta t = 50$ yr into (\ref{vtapprox}), we determine that nomadic worlds with $R \sim 150$ km may be reached by LEP.

Akin to NEP and LEP, solar electric propulsion (SEP) comprises a well-established avenue since the start of the 21st century \cite{RW02}. In SEP, solar arrays are deployed to collect power from the Sun, which is harnessed to produce electricity that is utilized for propulsion. A combination of SEP and radioisotope electric propulsion (REP) can permit the achievement of a terminal speed of $\sim 10$ AU/yr in the coming decades \cite{OD11,GM22}. On substituting $v_t \sim 10$ AU/yr and $\Delta t = 50$ yr into (\ref{vtapprox}), it is estimated that nomadic worlds with $R \sim 37$ km may be reached by SEP (allied to REP).

We reiterate that electric propulsion consists of numerous categories, owing to which our treatment has been selective; other pathways may warrant analysis.

\subsubsection{Nuclear fusion}\label{SSSecNucFus}
Hitherto, the propulsion systems covered arguably demonstrated working prototypes or the underlying physical principles were evinced (in)directly in other real-world systems. In contrast, after many decades, generating net energy from nuclear fusion and sustaining reactions for an extended time period remains deeply challenging, and essentially unrealized \cite{JPF08,FC16,BH16,OKW16}, although recent developments seem promising.\footnote{\url{https://www.llnl.gov/news/national-ignition-facility-achieves-fusion-ignition}} In fact, for the case of in-space power from aneutronic fusion, which bypasses the issues posed by generated neutrons, the 2015 NASA Technology Roadmaps assigned this technology an underwhelming status of TRL-1,\footnote{Refer to TA 3--65 in the roadmap: \url{https://www.nasa.gov/sites/default/files/atoms/files/2015_nasa_technology_roadmaps_ta_3_space_power_energy_storage.pdf}} while \cite{CCS15} proposed a slightly more optimistic estimate of TRL-2 for fusion propulsion in the near-future.

A valuable characteristic, however, of nuclear fusion is that markedly high exhaust speeds of $v_{ex} \sim 10^2$-$10^4$ km/s are viable in theory \cite[Section 5.3.6]{CCS15}. In view of the plethora of fusion propulsion schemes, we will restrict ourselves to only one proposal from the 21st century. The Direct Fusion Drive (DFD) is based on the field-reversed configuration \cite{GC02}, and could generate $v_{ex} \sim 225$ km/s (\cite{CS19}; see also \cite{TPC17,TPC18}). The DFD is reportedly well-suited for the exploration of the outer Solar system and reaching the SGL \cite{CS19,GG20}. Moreover, at least some of its auxiliary systems are rated at TRL-6 \cite{TPC17}.

For the DFD, on applying the scaling $v_t \lesssim 3 v_{ex}$ justified in Section \ref{SSSecEP}, we obtain $v_t \lesssim 142$ AU/yr. Finally, substituting this estimate into (\ref{vtapprox}) for $\Delta t = 50$ yr, we arrive at $R \lesssim 526$ km. This value is quite substantial, and is comparable to the radii of Tethys and Dione (moons of Saturn), Quaoar and Sedna (dwarf planets), and Ariel (a moon of Uranus), among other Solar system bodies. Even higher exhaust speeds may prove to be tenable for alternative versions of nuclear fusion, as mentioned earlier, thus permitting flybys of nomadic worlds with $R \gtrsim 10^3$ km. The full range of $v_t$ and $R$ for nuclear fusion propulsion is presented in Table \ref{TabPropSys}.

\begin{table*}
\begin{minipage}{165mm}
\caption{Approximate radius of nomadic worlds reachable by propulsion systems in $50$ years}
\label{TabPropSys}
\vspace{0.1 in}
\begin{tabular}{|c|c|c|}
\hline 
Propulsion system & Terminal speed (in AU/yr) & Radius (in km)\tabularnewline
\hline 
\hline 
Solar sails & $\sim 20$ & $\sim 75$ \tabularnewline
\hline 
Laser sails & $\lesssim 63$ to $\gtrsim 6.3 \times 10^3$ & $\lesssim 230$ to $\gtrsim 2.3 \times 10^4$\tabularnewline
\hline 
Magnetic sails & $\sim 20$ & $\sim 75$\tabularnewline
\hline 
Electric sails & $\sim 25$ & $\sim 93$ \tabularnewline
\hline 
Magnetoplasmadynamic thrusters & $\lesssim 63$ & $\lesssim 230$\tabularnewline
\hline 
Nuclear electric propulsion & $\sim 20$ & $\sim 75$\tabularnewline
\hline 
Laser electric propulsion & $\sim 40$ & $\sim 150$\tabularnewline
\hline 
Nuclear fusion & $\sim 63$ to $\sim 6.3 \times 10^3$ & $\sim 230$ to $\sim 2.3 \times 10^4$\tabularnewline
\hline 
Solar thermal propulsion & $\sim 20$ & $\sim 75$\tabularnewline
\hline 
Nuclear thermal propulsion & $\sim 20$ & $\sim 75$\tabularnewline
\hline 
\hline
Chemical propulsion & $\sim 7$ & $\sim 27$ \tabularnewline
\hline 
Solar electric propulsion & $\sim 10$ & $\sim 37$\tabularnewline
\hline 
\end{tabular}
\medskip

\vspace{0.1in}
\textbf{\textit{Additional notes:}} The last column indicates that nomadic worlds of radius $> R$ can be reached by a given propulsion system in flight time of $50$ years. It is estimated from (\ref{vtapprox}), after setting $\Delta t = 50$ yr and substituting the terminal speed (second column). All values should be viewed as strictly heuristic. The last two rows yield radii comfortably below $100$ km, implying that they might not be realistic options for reaching nomadic worlds in the stipulated timescale.
\end{minipage}
\end{table*}

\subsubsection{Solar and nuclear thermal propulsion}
Broadly speaking, solar thermal propulsion (STP) is predicated on executing a solar Oberth maneuver and harnessing the incident solar energy during this maneuver to heat the propellant. Initial studies suggested speeds of $\sim 20$ AU/yr are conceivable \cite{LEK01}, which is comparable to some of the propulsion systems examined hitherto. In the short term, achievable speeds may be restricted to $9$ AU/yr \cite{SPM21} owing to thermal issues \cite{MWG19}. However, in the long term, on substituting the above $v_t \sim 20$ AU/yr \cite{AA18,ASP20} and $\Delta t = 50$ yr in (\ref{vtapprox}), we arrive at $R \sim 75$ km.

In nuclear thermal propulsion (NTP), the energy released from nuclear fission of radioactive isotopes is used to heat cryogenic propellants, thereupon resulting in high specific impulses \cite{GH15}. Extensive tests since the 1970s \cite{JW91} have consequently identified a clear path for actualizing high TRL of $5$-$6$ for NTP \cite{HLH22}. Based on a recently proposed nuclear thermal rocket core \cite{YL22}, it has been estimated that speeds of $> 10$ AU/yr ought to be achievable \cite{HLH22}. Hence, over the span of a few decades, it is not inconceivable that $v_t \sim 20$ AU/yr would be technologically feasible, in which case we arrive at $R \sim 75$ km after invoking $\Delta t = 50$ yr and (\ref{vtapprox}).

\subsubsection{Is chemical propulsion tenable?}
Hitherto, we investigated a multitude of propulsion schemes. Furthermore, as delineated toward the start of Section \ref{SSecPropAnlys}, nuclear and antimatter propulsion are excluded, with the exception of fusion propulsion. Now, we will tackle chemical propulsion, which was not addressed so far.

A prominent and well-documented issue with chemical propulsion is that it evinces low exhaust velocities (albeit high thrust), implying that attaining high speeds (e.g., of order $100$ km/s) would necessitate impractically high amounts of fuel. However, this deficiency can be overcome to an extent through appropriate orbital maneuvers, of which the solar Oberth maneuver (SOM), to offer one example, is gaining traction \cite[e.g.,][]{HPE19,LS20,MWG22}.

By performing a detailed analysis, \cite{HLH22} explored the prospects for a mission to the putative Planet 9 at a potential distance of $\sim 450$ AU from the Sun using maneuvers such as SOM and the Jupiter Oberth maneuver (JOM) along with currently available chemical rockets. Interestingly, the authors found that objects at such distances approach the limits of chemical propulsion. More precisely, \cite{HLH22} determined that some of the trajectories lead to negative payloads, thereupon making them unfeasible. If we substitute this distance into (\ref{DISO}) and solve for the radius, we end up with $R \sim 34$ km, which is conspicuously lower than the size range employed in this work by a factor of $\gtrsim 3$.

The \emph{Interstellar Probe} concept is intended to survey the local interstellar medium (ISM) by means of deploying a combination of chemical propulsion and orbital maneuvers. It has been estimated that this mission can achieve final speeds of $\sim 7$-$8$ AU/yr \cite{BPC22}. If we adopt $v_t \sim 7.2$ AU/yr \cite{MWG22} and flight duration of $\Delta t = 50$ yr in (\ref{vtapprox}), we obtain $R \sim 27$ km, which is sufficiently close to the radius calculated in the above paragraph, thereby serving as a consistency check that is reported in Table \ref{TabPropSys}.

\subsection{Constraints on the flyby}\label{SSecConFly}
The chief goal of this paper is to assess the viability of various propulsion systems for reaching nomadic worlds in a stipulated timescale. We emphasize, however, that many other aspects will need to be evaluated for designing a full-fledged mission. We will briefly touch upon one of them below -- the constraints that must be fulfilled by a given flyby.

For starters, the spacecraft must fly close enough to the object so that it could be resolved (for surveying it with the spacecraft's instrumentation). If we assume a diffraction-limited telescope, then we require
\begin{equation}
    \frac{R}{D_s} \gtrsim \frac{\lambda}{\mathcal{D}},
\end{equation}
where $D_s$ is the distance between the spacecraft and the nomadic world during flyby, $\lambda$ is the wavelength, and $\mathcal{D}$ is the telescope aperture. For the telescope, we adopt $\lambda \sim 0.5$ $\mu$m and $\mathcal{D} \sim 0.2$ m, both of which are fiducial values chosen to match the LOng-Range Reconnaissance Imager (LORRI) on board the New Horizons spacecraft \cite{CWC08}. Hence, on rearranging the above expression, we obtain
\begin{equation}\label{DsLim}
    D_s \lesssim 2.7\,\mathrm{AU}\,\left(\frac{R}{1000\,\mathrm{km}}\right) \left(\frac{\mathcal{D}}{0.2\,\mathrm{m}}\right) \left(\frac{\lambda}{0.5\,\mathrm{\mu m}}\right)^{-1},
\end{equation}
indicating that the spacecraft must fly within a few AU of the nomadic world in order to resolve it. Note that the lower bound on $D_s$ can be boosted if a larger telescope is situated on board the spacecraft. Aside from this constraint on angular resolution, the telescope must intercept sufficient photons from the object to exceed its sensitivity limit(s), as it would be otherwise too dim to detect. Since this requirement involves additional parameters, which are subject to uncertainty, we shall not explicitly analyze it here.

In order to fulfill the condition (\ref{DsLim}), ideally we must have precise knowledge of the nomadic world's trajectory to plan the mission. The easiest scenario arises if the object were to be detected by an Earth-based telescope, along the lines described in Section \ref{SecDetect}, enabling us to continuously track it and estimate its trajectory accurately. In turn, a suitably accurate estimate of the latter would permit us, in principle, to design the spacecraft's trajectory such that it passes close enough to the nomadic world to meet the above criterion. However, since the nomadic world displays some uncertainty in its trajectory, the spacecraft's telescope may be deployed to detect it and make adjustments to the course. 

This approach was delineated in \cite[Section 3.1]{HEL22} for 1I/`Oumuamua, and is generalizable to nomadic worlds. On specifying $R \sim 1000$ km from earlier, we obtain $D_\mathrm{ISO} \sim 1.3 \times 10^4$ AU from (\ref{DISO}) and $H \approx 1.6$ after setting $\mathcal{A} \sim 0.1$ in (\ref{AbsMagInt}). We also select a relative speed of $\sim 100$ km/s based on previous sections (i.e., broadly the terminal velocity of the spacecraft), and adopt a worst-case fiducial positional uncertainty of $\theta_\mathrm{var} \sim 10^{-5}$ rad, which is commensurate with 1I/`Oumuamua \cite{TMH18,MFM18}, an ISO with unusual non-gravitational acceleration. This uncertainty might, however, be reduced significantly for generic nomadic worlds if their locations are precisely measured by Earth-based telescopes, as outlined toward the end of Section \ref{SecDetect} with respect to stellar occultations, and is discussed shortly below.

For this collection of parameters, inputting a desired apparent magnitude of $26$ for the object yields an integration time of $\sim 11$ h for a LORRI-type telescope, after using the calibrated data \cite{CWC08,LPW21}. The choice of this relatively faint limiting magnitude ensures that the object is discernible at adequate distance away from the spacecraft, so as to undertake appropriate maneuvers without expending too much $\Delta v$. From the chosen values of $H$ and apparent magnitude (\citep{KKO07} and \citep[Section 3.1]{HEL22}), we determine that the nomadic world could be detectable at a distance of roughly $\sim 5.7$ AU, which is equivalent to a time-of-flight of $\sim 10$ days for reaching the object. Hence, over this timescale, further maneuvers for accomplishing the flyby of the nomadic world can be effectuated; the maximal $\Delta v$ that might be warranted is $\sim \theta_\mathrm{var} D_\mathrm{ISO}$ (i.e., the distance uncertainty) divided by the corresponding time-of-flight.

Upon substituting the preceding parameters, we obtain $(\Delta v)_\mathrm{max} \sim 22$ km/s. In this worst case, substantial accelerations in arbitrary directions may prove to be necessary for performing a close flyby of the nomadic world. In this context, if, say, light sails or LEP/SEP are employed, the spacecraft would be far away from the power source (laser array or the Sun), thereby posing challenges to the actualization of the desired acceleration and spacecraft control. However, we reiterate and stress that the $\Delta v$ value is anticipated to decrease substantially if the positional uncertainty is orders of magnitude lower, as indicated earlier -- in fact, $\Delta v$ would decrease proportionally as per the above heuristic.

For instance, for a nomadic body with positional uncertainty restricted to $\theta_\mathrm{var} \sim 10^{-8}$ rad,\footnote{Similar accuracy has been achieved for several small bodies in the outer Solar system \cite[e.g.,][]{BAB11,PBP18,PSV} via astrometry; the Kuiper belt object 486958 Arrokoth, whose mean radius is $\sim 10$ km and semimajor axis is $\sim 45$ AU \cite{KPB22}, is one such example \cite{PBP18}. This uncertainty might be further improved by a large-scale occultation network \cite{RL19}.} we end up with $(\Delta v)_\mathrm{max} \sim 22$ m/s, after all other factors are held fixed. This $(\Delta v)_\mathrm{max}$ can be achieved, for example, by miniaturized electric propulsion systems powered by nuclear batteries \cite{hein2017andromeda,OHK21}. Field-emission electric propulsion (FEEP) could be one such viable technology \citep{BT18}. Even values of $(\Delta v)_\mathrm{max} \sim 220$ m/s, corresponding to $\theta_\mathrm{var} \sim 10^{-7}$ rad, are attainable by off-the-shelf deep space CubeSat propulsion systems (chemical, electric, or hybrid), as reviewed in \citep{cervone2022lumio}.

Hence, these constraints pertaining to the propulsion (sub)systems should be duly taken into account while designing the mission. For the worst case, it is conceivable that nuclear power in some fashion (e.g., NEP) might be harnessed -- either individually or jointly with the aforementioned propulsion systems -- since it can theoretically maintain constant power and controllability, thus alleviating the above issues. For substantially lower positional uncertainties, modest $\Delta v$ values are sufficient, which may be provided by cold gas or miniaturized electric propulsion powered by nuclear batteries. The suitable propulsion system(s) will be dictated, among other factors, by the positional uncertainty, the on-board instruments, power source(s), and so on, owing to which a generic analysis is not feasible.

Last, but not least, we have focused on flybys herein owing to the less stringent requirements, but future research is indubitably needed to assess the plausibility of sending spacecraft that could rendezvous with the nomadic world (e.g., orbiter and/or lander) and potentially initiate sample return, although the latter is bound to be extremely challenging in view of the distances and speeds involved.

\section{Discussion and Conclusions}\label{SecConc}
State-of-the-art observational data has demonstrated that interstellar space is populated by gravitationally unbound objects (i.e., nomads) spanning many orders of magnitude in size. Many compelling grounds exist for studying nomadic worlds, as outlined in Section \ref{SecIntro}. 

In particular, these objects could substantially advance our knowledge of astrophysics, planetary science, and astrobiology if they are located and then targeted by multiple missions (flybys followed by rendezvous and/or sample return missions). In this paper, taking our cue from this fundamental premise, we investigated near-future propulsion systems that can potentially enable us to reach (roughly spherical) nomadic worlds, with radii ranging from $\sim 100$ km to $\sim 10^4$ km, in $\lesssim 50$ years; such missions could offer a much higher scientific return than observations conducted by Earth- and space-based telescopes. To our knowledge, this paper is ostensibly the first in terms of assessing the feasibility of exploring a hidden continuum of objects between the stars (as opposed to traveling from one star to another), introducing an entire class of novel space missions geared towards unbounded celestial bodies.

We provided a heuristic estimate of the number density of nomadic worlds in Section \ref{SecDist}, and used this expression to gauge the characteristic distance to a nomadic world of radius $> R$. A couple of noteworthy conclusions arise from these calculations. First, the distance to the nearest nomadic ``planet'' might be several times smaller than the corresponding distance to the nearest known exoplanets, which orbit Proxima Centauri \cite[cf.][]{MF90,RT01}. Second, in the spherical volume bounded by Proxima Centauri, with Earth at the center, $\mathcal{O}(10)$ to $\mathcal{O}(100)$ (perhaps as high as $\sim 1000$) nomadic planets with sizes at least similar to that of Mercury may exist as per our simple prediction.

Next, we explored detectability of nomadic worlds in Section \ref{SecDetect}. We showed that larger worlds might be more amenable to discovery insofar as their thermal emission is concerned, even though they are fewer in number, and thus located at relatively greater distances. We briefly analyzed the prospects for discovering nomadic worlds by means of the Vera C. Rubin Observatory and the Hyper Suprime-Cam SSP Survey. Although the prospects for discovering nomads appear unlikely, we highlighted crucial caveats that could mitigate this result -- in this regard, we sketched the possibility of detecting nomadic worlds via alternative avenues such as stellar occultation.

Lastly, we motivated the ballistic assumption in Section \ref{SecProp}, and implemented this simplification to evaluate the viability of myriad propulsion systems for reaching nomadic worlds in a given timescale. Our salient findings are encapsulated in Table \ref{TabPropSys}. For the size range and flight time(s) we consider, chemical propulsion is apparently inadequate due to its low speeds, along expected lines. More accurately, chemical propulsion may allow us to probe objects with radius $\sim 30$ km, but this value falls comfortably below the lower bound prescribed earlier (namely, by a factor of about $3$). Nearly all the propulsion schemes tabulated in Table \ref{TabPropSys} could be deployed for not only exploring nearby nomadic worlds but also transporting payloads to the SGL at $\gtrsim 550$ AU \cite{TST20,FGT21,TGF23} in a few decades.

The majority of propulsion systems in Table \ref{TabPropSys}, such as electric sails and laser electric propulsion, permit us to reach nomadic worlds near the fiducial base limit of $\sim 100$ km. In other words, they are not well-suited for accessing truly planet-sized nomadic worlds of order $1000$ km. These worlds might be accessible to solar and electric sails launched in close proximity to the Sun, but these architectures are not tackled herein, as they may encounter additional obstacles (e.g., thermal stresses). Laser sails and nuclear fusion are the two conspicuous exceptions to this trend discernible from Table \ref{TabPropSys}. 

However, the chief drawback of nuclear fusion is that this technology has not been empirically demonstrated even in the case of sustained power generation on Earth, let alone spacecraft propulsion. Hence, it could be argued that laser sails constitute our best candidate for surveying nomadic worlds spanning an impressive range. In consequence, the continued development of laser sail missions such as \emph{Breakthrough Starshot} and its precursors augurs well for probing the planet-sized worlds closest to Earth (viz., nomadic planets) -- at potential distances of order $0.1$ pc -- and revolutionizing our understanding of diverse fields. 

It should be recognized that our work has addressed a vital engineering constraint, to wit, what propulsion systems would be appropriate for reaching a nomadic world (with radius $> R$) in a designated timescale. Yet, we caution that a number of other key engineering facets remain to be explored, such as optimal trajectories to nomadic worlds, as well as the design requirements, challenges, and solutions for enabling telecommunication and power storage. These issues are engendering active research on gram-sized laser sails \cite{LL21,PLB21,WGS21}, and analogous studies are needed for missions to nomadic worlds.

In summary, this paper delineated missions to nomadic worlds and elucidated their feasibility (in the near future) for the first time, to the best of our knowledge. We found that for several near-term propulsion systems, objects up to $\sim 100$ km in radius are within reach. For larger objects, only laser or fusion propulsion systems seem to be adequate, although the maturity of fusion propulsion is undeniably low. Laser sail propulsion might, however, see a clear path towards implementation with the \emph{Breakthrough Starshot} program. If this project or its variants come to fruition, it would lend credence to the optimistic notion that nomadic worlds of different sizes might be surveyed by spacecraft before the end of the 21st century, thereby opening a new era of planetary science -- the exploration of nomadic worlds traversing the outermost regions of the Solar System.

\section*{Acknowledgements}
The authors wish to thank Robert Kennedy, Jean Schneider, Madhur Tiwari, and the anonymous reviewers for valuable inputs concerning the paper. 

\bibliographystyle{elsarticle-num}
%\biboptions{authoryear}

\bibliography{RogueWorlds}

\end{document}